\documentclass[twocolumn,aps,prl,floatfix,superscriptaddress]{revtex4-1}

\usepackage{hyperref}
\usepackage[utf8]{inputenc}

\usepackage{graphicx}
\usepackage{dcolumn}
\usepackage{amsmath}
\usepackage{supertabular}
\usepackage{multirow}

\usepackage{color}
\usepackage{cancel}
\usepackage{ulem}
\usepackage{array}
\usepackage{tabularx}

\usepackage{amsmath}
\usepackage{amssymb}
\usepackage{xcolor}

\newcommand {\rootsNN}  {\ensuremath{\sqrt{s_{_{\rm{NN}}}}}}

\usepackage{multirow}

\begin{document}
%\linenumbers

\title{Onset of Constituent Quark Number Scaling in Heavy-Ion Collisions at RHIC} 

\bigskip

%%%% _______Author list______and Affiliation________

%_____________________________Author list and affilitions 

\affiliation{Academia Sinica, Taipei 115201}
\affiliation{Abilene Christian University, Abilene, Texas   79699}
\affiliation{AGH University of Krakow, FPACS, Cracow 30-059, Poland}
\affiliation{Argonne National Laboratory, Argonne, Illinois 60439}
\affiliation{American University in Cairo, New Cairo 11835, Egypt}
\affiliation{Ball State University, Muncie, Indiana 47306}
\affiliation{Brookhaven National Laboratory, Upton, New York 11973}
\affiliation{University of Calabria \& INFN-Cosenza, Rende 87036, Italy}
\affiliation{University of California, Berkeley, California 94720}
\affiliation{University of California, Davis, California 95616}
\affiliation{University of California, Los Angeles, California 90095}
\affiliation{University of California, Riverside, California 92521}
\affiliation{Central China Normal University, Wuhan, Hubei 430079}
\affiliation{University of Illinois at Chicago, Chicago, Illinois 60607}
\affiliation{Chongqing University, Chongqing, 401331}
\affiliation{Creighton University, Omaha, Nebraska 68178}
\affiliation{Czech Technical University in Prague, FNSPE, Prague 115 19, Czech Republic}
\affiliation{Technische Universit\"at Darmstadt, Darmstadt 64289, Germany}
\affiliation{National Institute of Technology Durgapur, Durgapur - 713209, India}
\affiliation{ELTE E\"otv\"os Lor\'and University, Budapest H-1117, Hungary}
\affiliation{Frankfurt Institute for Advanced Studies FIAS, Frankfurt 60438, Germany}
\affiliation{Fudan University, Shanghai, 200433 }
\affiliation{Guangxi Normal University, Guilin, 541004}
\affiliation{University of Heidelberg, Heidelberg 69120, Germany }
\affiliation{University of Houston, Houston, Texas 77204}
\affiliation{Huzhou University, Huzhou, Zhejiang  313000}
\affiliation{Indian Institute of Science Education and Research (IISER), Berhampur 760010 , India}
\affiliation{Indian Institute of Science Education and Research (IISER) Tirupati, Tirupati 517507, India}
\affiliation{Indian Institute Technology, Patna, Bihar 801106, India}
\affiliation{Indiana University, Bloomington, Indiana 47408}
\affiliation{Institute of Modern Physics, Chinese Academy of Sciences, Lanzhou, Gansu 730000 }
\affiliation{University of Jammu, Jammu 180001, India}
\affiliation{Kent State University, Kent, Ohio 44242}
\affiliation{University of Kentucky, Lexington, Kentucky 40506-0055}
\affiliation{Lanzhou University, Lanzhou 730000}
\affiliation{Lawrence Berkeley National Laboratory, Berkeley, California 94720}
\affiliation{Lehigh University, Bethlehem, Pennsylvania 18015}
\affiliation{Max-Planck-Institut f\"ur Physik, Munich 80805, Germany}
\affiliation{Michigan State University, East Lansing, Michigan 48824}
\affiliation{National Institute of Science Education and Research, HBNI, Jatni 752050, India}
\affiliation{National Cheng Kung University, Tainan 70101}
\affiliation{Nuclear Physics Institute of the CAS, Rez 250 68, Czech Republic}
\affiliation{The Ohio State University, Columbus, Ohio 43210}
\affiliation{Panjab University, Chandigarh 160014, India}
\affiliation{Purdue University, West Lafayette, Indiana 47907}
\affiliation{Rice University, Houston, Texas 77251}
\affiliation{Rutgers University, Piscataway, New Jersey 08854}
\affiliation{University of Science and Technology of China, Hefei, Anhui 230026}
\affiliation{South China Normal University, Guangzhou, Guangdong 510631}
\affiliation{Sejong University, Seoul, 05006, South Korea}
\affiliation{Shandong University, Qingdao, Shandong 266237}
\affiliation{Shanghai Institute of Applied Physics, Chinese Academy of Sciences, Shanghai 201800}
\affiliation{Southern Connecticut State University, New Haven, Connecticut 06515}
\affiliation{State University of New York, Stony Brook, New York 11794}
\affiliation{Instituto de Alta Investigaci\'on, Universidad de Tarapac\'a, Arica 1000000, Chile}
\affiliation{Temple University, Philadelphia, Pennsylvania 19122}
\affiliation{Texas A\&M University, College Station, Texas 77843}
\affiliation{Texas Southern University, Houston, Texas 77004}
\affiliation{University of Texas, Austin, Texas 78712}
\affiliation{Tsinghua University, Beijing 100084}
\affiliation{University of Tsukuba, Tsukuba, Ibaraki 305-8571, Japan}
\affiliation{University of Chinese Academy of Sciences, Beijing, 101408}
\affiliation{United States Naval Academy, Annapolis, Maryland 21402}
\affiliation{Valparaiso University, Valparaiso, Indiana 46383}
\affiliation{Variable Energy Cyclotron Centre, Kolkata 700064, India}
\affiliation{Warsaw University of Technology, Warsaw 00-661, Poland}
\affiliation{Wayne State University, Detroit, Michigan 48201}
\affiliation{Wuhan University of Science and Technology, Wuhan, Hubei 430065}
\affiliation{Yale University, New Haven, Connecticut 06520}

\author{B.~E.~Aboona}\affiliation{Texas A\&M University, College Station, Texas 77843}
\author{J.~Adam}\affiliation{Czech Technical University in Prague, FNSPE, Prague 115 19, Czech Republic}
\author{L.~Adamczyk}\affiliation{AGH University of Krakow, FPACS, Cracow 30-059, Poland}
\author{I.~Aggarwal}\affiliation{Panjab University, Chandigarh 160014, India}
\author{M.~M.~Aggarwal}\affiliation{Panjab University, Chandigarh 160014, India}
\author{Z.~Ahammed}\affiliation{Variable Energy Cyclotron Centre, Kolkata 700064, India}
\author{A.~K.~Alshammri}\affiliation{Kent State University, Kent, Ohio 44242}
\author{E.~C.~Aschenauer}\affiliation{Brookhaven National Laboratory, Upton, New York 11973}
\author{S.~Aslam}\affiliation{Fudan University, Shanghai, 200433 }
\author{J.~Atchison}\affiliation{Abilene Christian University, Abilene, Texas   79699}
% ...

\collaboration{STAR Collaboration}\noaffiliation

\date{\today}
\begin{abstract}

Partonic collectivity is one of the necessary signatures for the formation of quark-gluon plasma in high-energy nuclear collisions. Number of constituent quarks (NCQ) scaling has been observed for hadron elliptic flow $v_2$ in top energy nuclear collisions at the Relativistic Heavy Ion Collider and the LHC, 
and this has been theoretically suggested as strong evidence for partonic collectivity. 
In this Letter, a systematic analysis of $v_2$ of $\pi^{\pm}$, $K^{\pm}$, $K^{0}_{S}$, $p$, and $\Lambda$ in Au+Au collisions at ${\rootsNN}$ = 3.2, 3.5, 3.9, and 4.5 GeV,  with the STAR experiment at the Relativistic Heavy Ion Collider, is presented. 
NCQ scaling is markedly violated at 3.2 GeV, consistent with a hadronic-interaction dominated equation of state.
However, as the collision energy increases, a gradual evolution to NCQ scaling is observed.  
This beam-energy dependence of $v_2$ for all hadrons studied provides evidence for the onset of dominant partonic interactions by ${\rootsNN}$ = 4.5 GeV. 

\end{abstract}

\maketitle
%\nolinenumbers
%%%%%%%%%%%%%%%%%%  Main article  %%%%%%%%%%%%%%%%%%
Elliptic flow ($v_2$), the second-order harmonic coefficient in the Fourier expansion of the final state particle azimuthal distribution 
with respect to the reaction plane, is sensitive to constituent interactions and the degrees of freedom of 
the matter created in heavy-ion collisions~\cite{Voloshin:2008dg}. 
The significant $v_2$ signal and the number of constituent quarks (NCQ) scaling of this elliptic flow 
are considered as evidence of quark-gluon plasma (QGP)
formation in high-energy relativistic heavy-ion collisions~\cite{STAR:2005gfr, PHENIX:2004vcz, STAR:2015gge, STAR:2017kkh, Braun-Munzinger:2007edi}.

NCQ scaling is understood within the coalescence picture, where hadrons are formed through quark recombination~\cite{Molnar:2003ff}. 
In this framework, light quarks ($u$, $d$, $s$) are assumed to exhibit collective behavior consistent with hydrodynamic flow, 
leading to a universal pattern of hadron $v_2$ when scaled by the number of constituent quarks, 
indicating the presence of quark degrees of freedom in the medium. 
As the collision energy gradually decreases below a certain threshold, the high temperature and energy density conditions necessary for the formation of QGP will no longer be satisfied. 
%Consequently, experimental signals of QGP such as NCQ scaling are expected to disappear.
Consequently, the NCQ scaling of elliptic flow is expected to disappear.
The Beam Energy Scan (BES) program at the Relativistic Heavy Ion Collider at Brookhaven National Laboratory (RHIC) aims to explore the quantum chromodynamics (QCD) phase structure by lowering the collision energy, 
spanning an energy range from {\rootsNN} = 3 to 62.4 GeV,
in search of possible signals for a QCD first-order phase boundary and a critical point 
through heavy-ion collision experiments~\cite{phase_tran:2010, Bzdak:2019pkr, Luo:2020pef, Chen:2024vb}. 

In the elliptic flow measurements of the first phase of the RHIC BES, we observed a relatively good agreement of NCQ scaling in collisions with {\rootsNN} $\geq$ 7.7 GeV~\cite{STAR:2013cow, STAR:2013ayu, STAR:2015rxv, STAR:2012och}, 
although observations of a possible deviation from NCQ scaling, around 2$\sigma$, were noted for the $\phi$ meson $v_2$ in collisions at {\rootsNN} = 7.7 and 11.5 GeV~\cite{STAR:2013cow, STAR:2013ayu, STAR:2015rxv, STAR:2012och}. Further investigation with larger data samples is warranted. 
The latest published elliptic flow results from the STAR experiment at {\rootsNN} = 3 GeV show that NCQ scaling breaks among $\pi^+$, $K^+$, and $p$ at this energy~\cite{STAR:2021yiu}.

The second phase of the RHIC BES (BES-II) focuses on energies ranging from {\rootsNN} = 3 to 19.6 GeV, corresponding to a baryon chemical potential range of 750 to 205 MeV~\cite{Cleymans:2005xv, Andronic:2005yp, STAR:2017sal}. 
To do so, STAR has conducted a series of detector upgrades for BES-II, including 
an inner time projection chamber (iTPC) to improve the track quality~\cite{Yang:2017llt};
an event plane detector (EPD) to measure the collision centrality and the event plane of the collision event~\cite{Adams:2019fpo};
an endcap time of flight (eTOF) to enhance the particle identification capability 
in the midrapidity region for the fixed-target (FXT) program~\cite{Wang:2023vqi}.

\begin{figure}[htbp]
  %\centering
  \includegraphics[width=0.96\columnwidth]{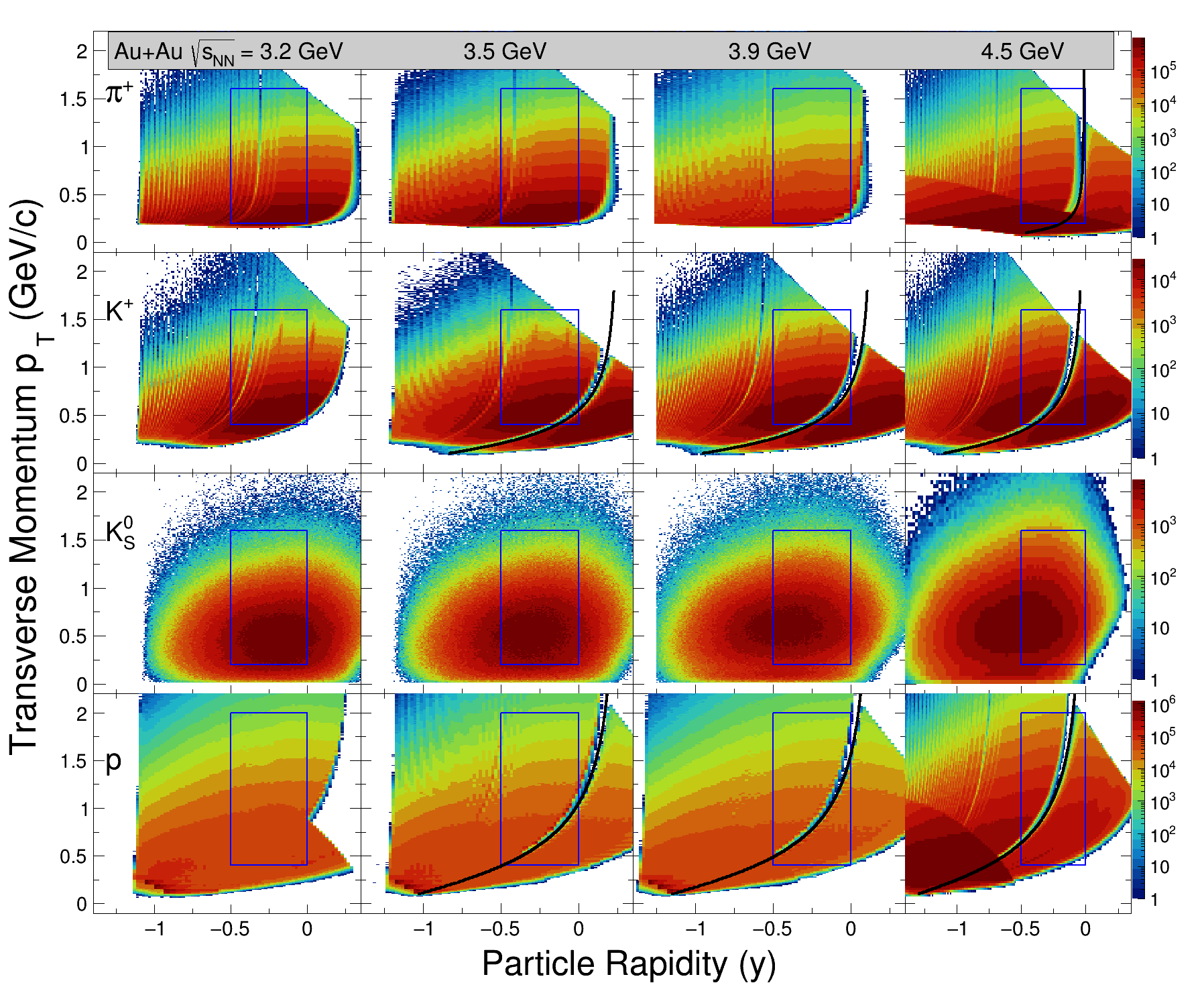}
  \caption{The transverse momentum ($p_T$) and identified particle rapidity ($y$, in the center-of-mass frame) distributions 
  for $\pi^{+}$, $K^{+}$, $K^{0}_{S}$, and $p$ from Au+Au collisions at {\rootsNN} = 3.2, 3.5, 3.9, and 4.5 GeV.
  To the right of the black curve is the acceptance provided by eTOF.
  The blue boxes represent the acceptance ($-0.5 < y < 0 $) used for elliptic flow measurements.}     
  \label{Fig:fig1}
\end{figure}

\begin{figure*}[!htb]
%\begin{figure}[!htb]
    \centering
    \includegraphics[width=0.99\textwidth]{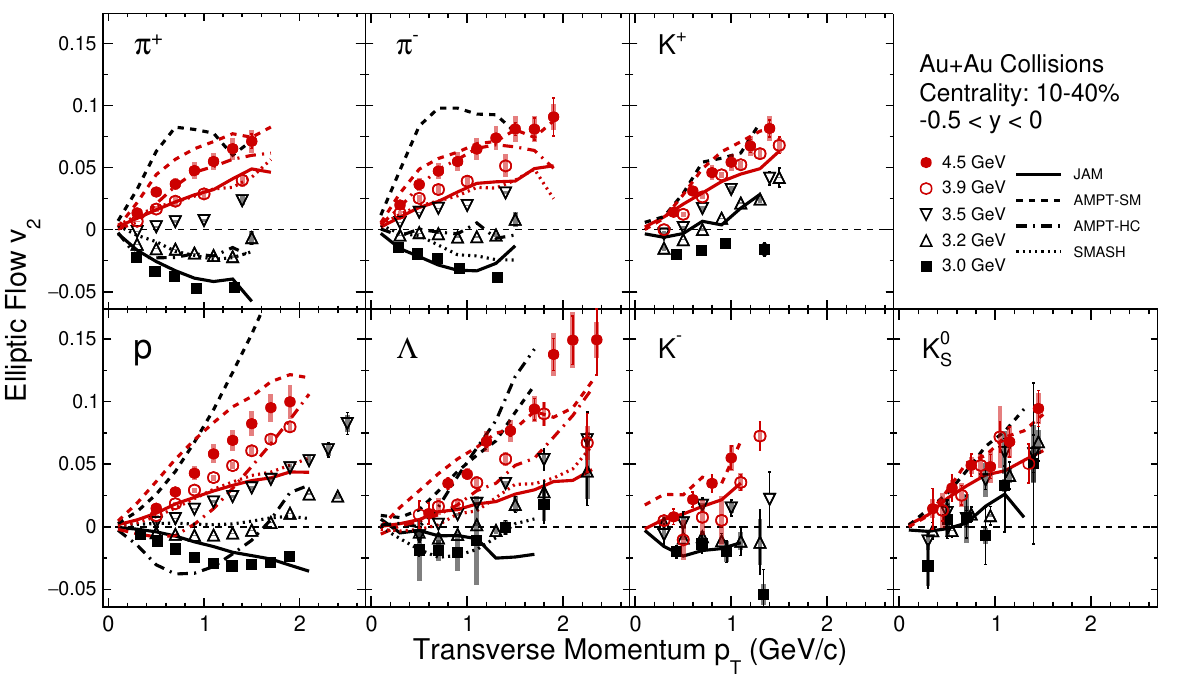}
    \caption{Transverse momentum ($p_T$) dependence of $v_2$ for 
    $\pi^{\pm}$, $K^{\pm}$, $K^{0}_{S}$, $p$, and $\Lambda$ in 10\%--40\% centrality 
    for Au+Au collisions at {\rootsNN} = 3.0, 3.2, 3.5, 3.9, and 4.5 GeV.
    Statistical and systematic uncertainties are shown as bars and bands, respectively. 
    Different curves represent the results from  JAM (solid), AMPT-SM (dashed), 
    AMPT-HC (dash-dotted), and SMASH (dotted) calculations: red for 4.5 GeV, and black for 3.0 GeV.
    For clarity, the uncertainties of the model calculations (relative error $\le$ 10\%) are not shown. 
    Calculations of SMASH and AMPT-HC for $K^{\pm}$, $K^{0}_{S}$ are omitted, 
    as the results are too similar to JAM to be clearly distinguished. Additionally, the $K^{-}$ results at 3.0 GeV are not presented due to the low production yield in the model.}
    \label{Fig:fig2}
%\end{figure}
\end{figure*}

In this Letter, we report $v_2$ measurements for 
$\pi^{\pm}$, $K^{\pm}$, $K^{0}_{S}$, $p$, and $\Lambda$ 
in Au+Au collisions at {\rootsNN} = 3.2, 3.5, 3.9, and 4.5 GeV. 
The data were collected during the 2019 and 2020 runs of the STAR FXT program at RHIC, 
with 220, 100, 100, and 120 $\times \ 10^{6}$ minimum-bias events at 3.2, 3.5, 3.9, and 4.5 GeV, 
respectively. 
Datasets for collision energies above 4.5 GeV in FXT mode are not included due to limited midrapidity coverage.
The primary vertex position of each event along the beam direction 
is selected to be within 2 cm from the target, which was located 200 cm from the center of the time projection chamber (TPC).
Additionally, the vertex location along the radial direction is chosen to be smaller than 2 cm to eliminate possible beam interactions with the vacuum pipe.
Runs where the mean value of physics variables, such as azimuthal angle and pseudorapidity, 
deviates from the overall mean by more than 5 standard deviations are labeled as bad runs and excluded from the analysis~\cite{STAR:2021mii}.
Pileup events, resulting from the long drift time of TPC electrons relative to the time interval between beam bunches, leading to the misidentification of multiple events as a single event,
are removed by correlating the TPC multiplicity with the time of flight (TOF) matched multiplicity.
Collision centralities are determined by fitting the TPC-measured charged particle multiplicity 
with a Monte Carlo Glauber model~\cite{Miller:2007ri}, 
where 0\% indicates full nuclear overlap (most central collisions).
To select high-quality tracks, we require a distance of closest approach from the vertex  $\leq$ 3 cm and a minimum of 15 space points within the acceptance of the TPC.
For particle identification of $\pi^{\pm}, K^{\pm},$ and $p$, a combination of the TPC and the TOF detector is used, 
which relies on the ionization energy loss information and time-of-flight information, respectively. 
A minimum identification purity of $>$ 90\% is required for elliptic flow measurements, with the particle identification contamination effect estimated as a systematic uncertainty source.
The strange hadrons $K_S^0$ and $\Lambda$ are reconstructed by pairing their daughter tracks via the Kalman filter particle package~\cite{Kisel:2018nvd, Banerjee:2020iab}. 
%This package also utilizes the covariance matrix of reconstructed tracks to construct a set of topological variables.

\begin{figure*}[htbp]
    \centering
    \includegraphics[width=0.99\textwidth]{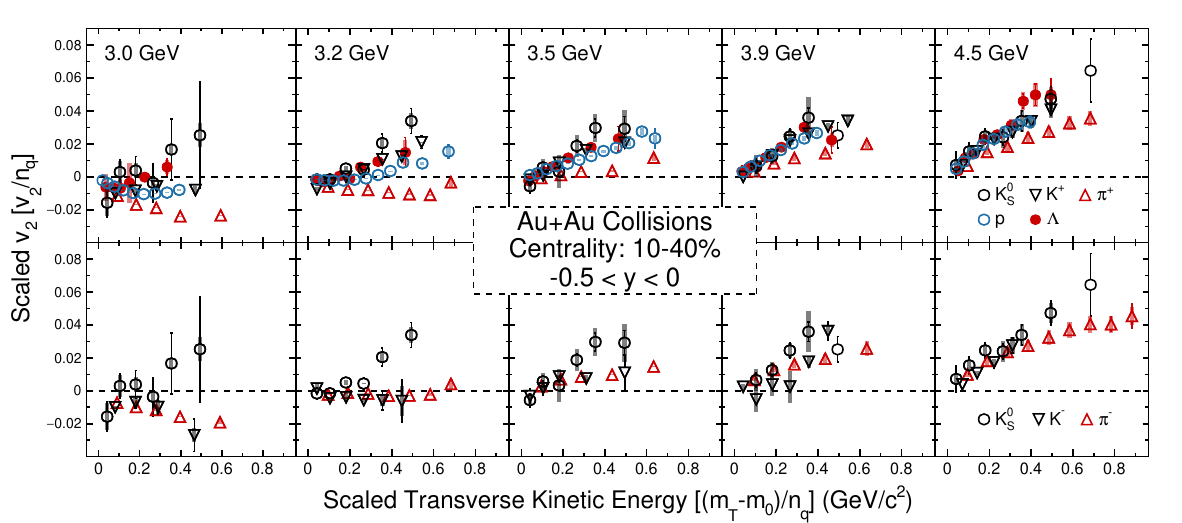}
    \caption{The number of constituent quarks $n_q$ scaled $v_2$ as a function of $n_q$ scaled transverse kinetic energy for particles (upper panel) and antiparticles (lower panel) 
    in 10\%--40\% centrality for Au+Au collisions at {\rootsNN} = 3.0, 3.2, 3.5, 3.9, and 4.5 GeV. 
    Statistical and systematic uncertainties are shown as bars and bands, respectively.}
    \label{Fig:fig3}
\end{figure*}

\begin{figure}[htb]
    \centering
    \includegraphics[width=0.47\textwidth]{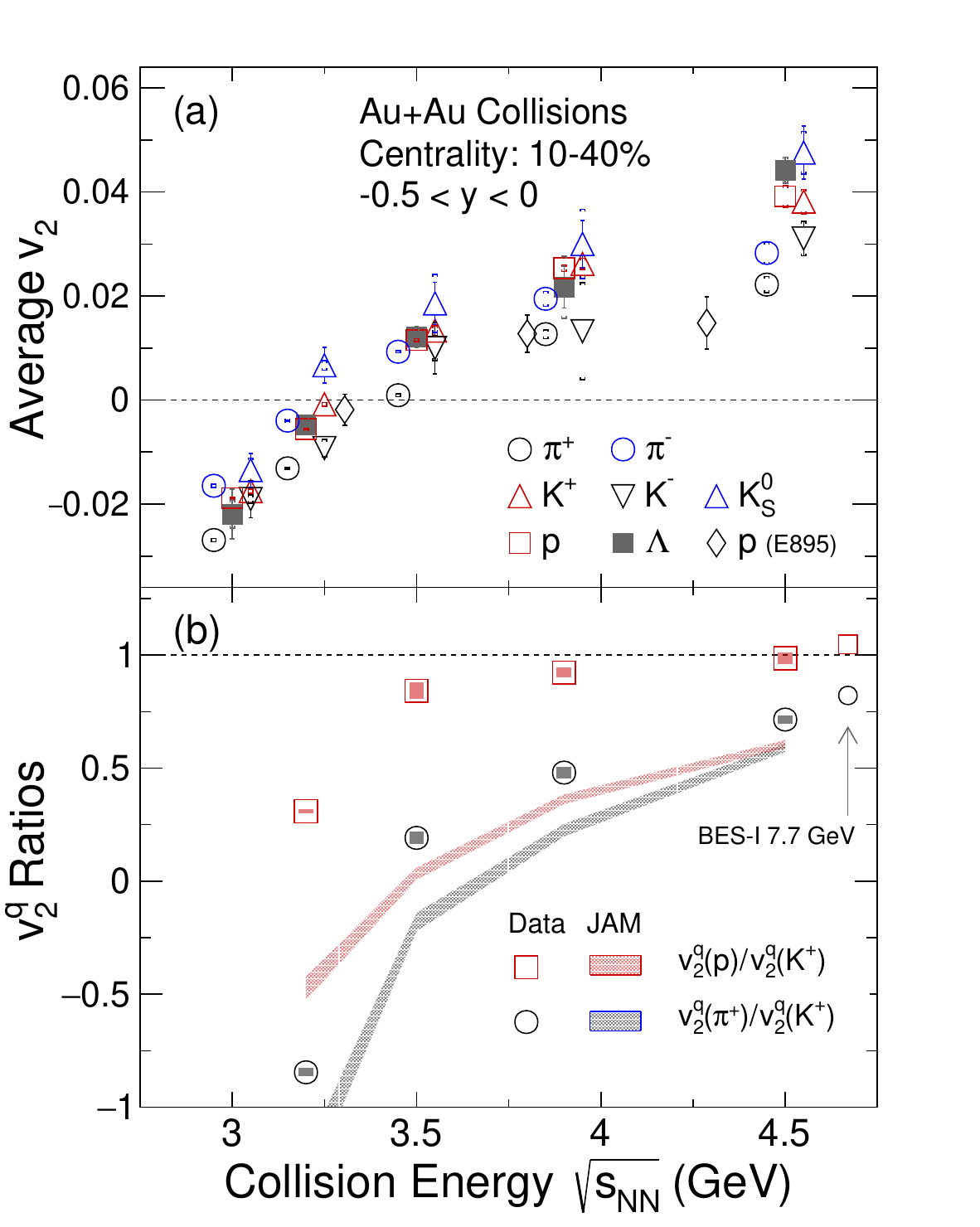}
    \caption{(a) Energy dependence of $p_T$ integrated $v_2$ for $\pi^{\pm}$, $K^{\pm}$, $K^{0}_{S}$, $p$, and $\Lambda$  
    in 10\%--40\% centrality from Au+Au collisions at {\rootsNN} = 3.0, 3.2, 3.5, 3.9, and 4.5 GeV. 
    For clarity, the x axis values of pions and kaons are shifted by $\pm 0.05$ respectively. 
    Proton data of E895 collaboration are from Refs.~\cite{E895:1999ldn, Herrmann:1999wu}.
    (b) Energy dependence of $n_{q}$ scaled $v_{2}$ ratios of $v_{2}^{q}(\pi^{+}) / v_{2}^{q}(K^{+})$ and $v_{2}^{q}(p) / v_{2}^{q}(K^{+})$ at ($m_T - m_0$)/$n_q$ = 0.4 GeV/$c^2$ in the same centrality. The ratios at 7.7 GeV are calculated based on the data in Ref.~\cite{STAR:2013ayu}.
    Statistical and systematic uncertainties are shown as bars and bands, respectively. 
    The JAM calculations with baryonic mean field are shown as dashed bands.}%: grey for $\pi^{+} / K^{+}$, red for $p / K^{+}$.}
    \label{Fig:fig4}
\end{figure}

The transverse momentum ($p_T$) and rapidity ($y$, in the center-of-mass frame) distributions of identified particles $\pi^{+}, K^{+}, K^{0}_{S},$ and $p$ 
from Au+Au collisions at {\rootsNN} = 3.2, 3.5, 3.9, and 4.5 GeV are shown in Fig.~\ref{Fig:fig1}. 
The additional acceptance provided by the eTOF (right side of the black curve) is particularly evident at forward rapidities.
The blue boxes show the region ($-0.5 < y < 0 $) used to determine the elliptic flow measurements reported in this Letter.
Because of the asymmetry of the phase space acceptance in fixed-target collisions, the three-sub event method 
is applied to reconstruct the event plane and estimate the event plane resolution~\cite{Poskanzer:1998yz}:
\begin{equation}
R(a)=\sqrt{\frac{\left\langle\cos \left[n\left(\Psi_m^a-\Psi_m^b\right)\right]\right\rangle\left\langle\cos \left[n\left(\Psi_m^a-\Psi_m^c\right)\right]\right\rangle}{\left\langle\cos \left[n\left(\Psi_m^b-\Psi_m^c\right)\right]\right\rangle}},
\end{equation}
where $n$ denotes the corresponding Fourier coefficient $v_n$, 
and $m$ indicates the $m$ th order harmonic event plane. 
$\Psi_{m}^{a}$, $\Psi_{m}^{b}$, and $\Psi_{m}^{c}$ represent the three-sub event planes, 
determined with pseudorapidity ($\eta_{\text{Lab}}$) ranges of [3.2, 6.1], [2.8, 3.1], and [0, 1.0], 
respectively, in the laboratory frame.
%As the small $v_{2}$ signal in the forward rapidity region, 
The first-order coefficient ($v_1$) is more significant than $v_2$ within this energy region~\cite{HADES:2020lob, STAR:2017sal}, 
and $\psi_1$ and $\psi_2$ are strongly correlated.
Therefore, $v_2$ is measured with respect to the first-order event plane, 
with event plane resolution about 20\%--40\% in mid-central 10\%--40\% collisions. 
Possible contributions from rapidity-even $v_1$ due to dipolar fluctuations 
are expected to be subdominant at these collision energies.

The $p_{T}$ dependence of $v_{2}$ measurements considers the detector efficiency as a function of transverse momentum $p_{T}$ and rapidity $y$. 
This efficiency encompasses the track reconstruction efficiency of the TPC and the TOF matching efficiency for $\pi^{\pm}$, $K^{\pm}$, and $p$, 
as well as the additional reconstruction efficiency for $K^{0}_{S}$ and $\Lambda$. 
These efficiencies are estimated using the embedding method within the STAR analysis framework~\cite{STAR:2008med, STAR:2010yyv, STAR:2017sal, STAR:2021iop}.

The systematic uncertainties in the measurements are determined by varying the analysis cuts mentioned above, 
which include track quality cuts, particle identification cuts, and event plane resolution. 
For each cut variable, we assign the maximum deviation from the default value as the systematic error originating from that source. 
Assuming these sources are uncorrelated, the total systematic uncertainty is calculated by summing them together quadratically.
A Barlow check~\cite{Barlow:2002yb} was implemented to eliminate the uncertainties introduced by statistical fluctuations during this process.
The largest systematic uncertainty in proton $v_{2}$ at 4.5 GeV, arising from event plane resolution, is less than 13.3\%. 
The systematic uncertainty from particle identification cuts is less than 1.5\%, and less than 1.7\% for track quality cuts.

Figure \ref{Fig:fig2} presents the $p_T$ dependence of $v_2$ for $\pi^{\pm}$, $K^{\pm}$, $K^{0}_{S}$, $p$, and 
$\Lambda$ in 10\%--40\% centrality for Au+Au collisions at {\rootsNN} = 3.0, 3.2, 3.5, 3.9, and 4.5 GeV. 
The data at 3.2, 3.5, 3.9, and 4.5 GeV represent new measurements, while the 3.0 GeV data are taken from a previous publication~\cite{STAR:2021yiu}.
Because of the rarity of $\bar{p}$ and $\bar{\Lambda}$ in this collision energy range, 
measurements of the elliptic flow for these two particles are not available.
At lower collision energies, the passing time of projectile and target spectators 
is comparable to the mean freeze-out time of produced particles~\cite{Liu:1998yc, LeFevre:2016vpp, Liu:2024ugr}. 
As a result, in-plane expansion is hindered by the spectators, leading to negative $v_2$ values~\cite{HADES:2020lob, HADES:2022osk}. 
This effect, known as spectator shadowing, is particularly evident at 3.0 GeV, 
where hadrons dominate the medium's degrees of freedom, 
characterized by large cross sections and short mean-free paths. 
However, as the collision energy increases, the suppression effect weakens, 
and $v_2$ transitions from negative to positive between 3.0 and 4.5 GeV, as shown in Fig. \ref{Fig:fig2}. 
This trend suggests the gradual emergence of partonic degrees of freedom, 
where interactions are governed by smaller partonic cross sections and longer mean-free paths.

\iffalse
Clear energy dependence of $v_2$ is observed for each particle species.
In lower collision energies, the passing time ($\sim 2R/\gamma \beta$) of the projectile and target spectators 
is comparable to the mean time of particle freeze-out~\cite{Liu:1998yc, LeFevre:2016vpp, Liu:2024ugr}, 
where $R$ represent the radius of nuclear size, $\beta$ denotes velocity, 
$\gamma$ indicates Lorentz contraction factor. 
As a result, the in-plane expansion is hindered by the spectators, a phenomenon known as the shadowing effect. 
Particles are preferentially emitted in the direction perpendicular to the reaction plane, leading to a negative $v_2$ signal~\cite{HADES:2020lob, HADES:2022osk}.
The $v_2$ as a function of $p_{T}$ changes from negative to positive between 3.0 GeV and 4.5 GeV, 
indicating that the spectator-shadowing effect decreases rapidly within this energy range. 
\fi

Comparisons with transport models further support this interpretation.
The calculations from the Jet AA Microscopic Transport Model (JAM)~\cite{Nara:2021fuu, Nara:2022kbb}, 
Multi-Phase Transport Model: Hadron Cascade (AMPT-HC) and String Melting (AMPT-SM) mode~\cite{PhysRevC.72.064901, Yong:2023uct}, 
and Simulating Many Accelerated Strongly interacting Hadrons (SMASH)~\cite{SMASH:2016zqf} are represented by the curves. 
For the lowest collision energy 3.0 GeV,  
the hadronic transport models JAM, AMPT-HC, and SMASH qualitatively describe the $v_2$ data.
The multiphase transport model AMPT-SM (black dashed curve) predicts the opposite sign of $v_2$
because it does not account for the finite thickness of the incoming nuclei, 
thus missing the potential shadowing effect from spectator nucleons.
%which could be due to the spectator-shadowing effect is not properly taken into account.
For 4.5 GeV, the hadronic transport models generally underestimate the $v_2$ data (except $\pi^{\pm}$ from AMPT-HC); 
in contrast, AMPT-SM, which incorporates partonic interactions, better describes the $v_2$ data.
This suggests that parton interactions and the coalescence play an important role in generating such a significant $v_2$ signal.
Detailed comparison of the model calculations can be found in Supplemental Material~\cite{2qhx-cp79}.

NCQ scaling is expected to reflect the effective degrees of freedom of the medium.
The elliptic flow scaled by the number of constituent quarks, $v_{2}/n_{q}$, 
as a function of the scaled transverse kinetic energy ($m_T - m_0$)/$n_q$ 
for particles (upper panel) and antiparticles (lower panel) separately in 10\%--40\% 
centrality for Au+Au collisions is shown in Fig.~\ref{Fig:fig3}, 
where $m_T = \sqrt{(p_T^2 + m_0^2)}$, $m_0$ is the rest mass of the particle.
In collisions at 3.0 and 3.2 GeV, it can be clearly observed that the NCQ scaling is broken, with each particle exhibiting a different trend.
As the collision energy increases from 3.2 to 4.5 GeV, the NCQ scaling becomes evident. 
These observations suggest that hadronic interactions dominate the equation of state of the created matter at 3.0 and 3.2 GeV, 
while partonic interactions become more important at collision energies greater than 3.2 GeV.
On the model side, 
JAM better describes the NCQ breaking at 3.0 GeV but fails to capture the scaling behavior at 4.5 GeV;
AMPT-SM shows better scaling behavior than hadronic transport models at 4.5 GeV. 
Notably, the $v_{2}$ of $\pi^{\pm}$ is smaller than the scaling of other particles at 4.5 GeV. 
The $p_{T}/n_{q}$ scaling exhibits better performance than $(m_T - m_0)/n_q$ for $\pi^{\pm}$, 
suggesting that the deviation at this energy
is primarily attributed to the significantly smaller mass of pions compared to other hadrons.
%This finding is also consistent with the predictions from the original papers that proposed $p_T$-based scaling
 It is noteworthy that the original NCQ scaling was proposed based on $p_T$~\cite{Molnar:2003ff, Voloshin:2005up}.
Detailed model comparisons and scaling results are given in Supplemental Material~\cite{2qhx-cp79}.

We further investigate the $p_T$ integrated $v_2$ as a function of collision energy.
Figure ~\ref{Fig:fig4} (a) shows the energy dependence of $p_T$ integrated $v_2$ for $\pi^{\pm}$, 
$K^{\pm}$, $K^{0}_{S}$, $p$, and $\Lambda$
in 10\%--40\% centrality from Au+Au collisions at {\rootsNN} = 3.0, 3.2, 3.5, 3.9, and 4.5 GeV. 
The integrated $v_2$ is calculated within $0.2 < p_{T} (\text{GeV}/c) < 1.6$ for $\pi^{\pm}$, 
$0.4 < p_{T} (\text{GeV}/c) < 1.6$ for  $K^{\pm}, K^{0}_{S}$, $0.4 < p_{T} (\text{GeV}/c) < 2.0$ for $p, \Lambda$.
$p_T$ integrated $v_2$ changes from negative to positive as the energy increases from 3.0 to 4.5 GeV, crossing zero at about 3.2 GeV.
Compared to previously published results from the E895 Collaboration~\cite{E895:1999ldn, Herrmann:1999wu}, 
our data show good agreement near the zero-crossing point. At higher energies, however, 
our measurements yield larger $v_2$ values, which may be attributed to the use of midcentral (10\%--40\%) collisions, 
whereas the E895 results correspond to collisions with an impact parameter of 5--7 fm (approximately 10--15\% centrality).
Clear differences between $\pi^{-}$ and $\pi^{+}$ are observed at each energy, and the differences become smaller as the energy increases. 
This is consistent with the picture of the baryon number transport--quarks transported from beam rapidity to midrapidity experience more violent scatterings than quarks produced at midrapidity~\cite{Dunlop:2011cf}. 
Additionally, the initial nuclear matter is a neutron-rich environment, causing a larger transported effect for $\pi^{-} (\bar{u}d)$ compared to $\pi^{+}(u\bar{d})$.
Although the uncertainties are large for $K^{\pm}, K^{0}_{S}$, these three kaons exhibit ordering behavior, i.e., 
$K^{0}_{S}(d\bar{s}) > K^{+}(u\bar{s}) > K^{-}(\bar{u}s) $, which is also consistent with the transported effect. 
These observations indicate the effect of the transported quarks 
is also present even at this energy region where the baryon density is high.
On the other side, the $v_2$ of $p$ and $\Lambda$ are consistent within statistical uncertainties.

In order to quantify the trend of NCQ scaling with collision energy, Fig.~\ref{Fig:fig4} (b) shows 
the $n_q$ scaled $v_2$ ratios of $v_{2}^{q}(\pi^{+}) / v_{2}^{q}(K^{+})$ and $v_{2}^{q}(p) / v_{2}^{q}(K^{+})$ 
at ($m_T - m_0$)/$n_q$ = 0.4 GeV/$c^2$ as a function of collision energy,
where the $v_{2}^{q}$ represents the $n_{q}$ scaled $v_2$ ($v_{2}/n_{q}$).
The ratio of $v_{2}^{q}(p) / v_{2}^{q}(K^{+})$ is close to unity at 3.9 and 4.5 GeV, while it deviates significantly at 3.2 GeV. 
Although hadronic model JAM calculations fit the $v_{2} (p_{T})$ data better at lower collision energies, they underestimate the ratios throughout the energy range studied.

In summary, we present the elliptic flow of identified hadrons $\pi^{\pm}$, $K^{\pm}$, $K^{0}_{S}$, $p$, and $\Lambda$
in Au+Au collisions at {\rootsNN} = 3.2, 3.5, 3.9, and 4.5 GeV. 
The $v_2$ of these particles changes from negative to positive around 3.2 GeV. At the lower colliding energy, {\rootsNN} $\leq$ 3.2 GeV, 
NCQ scaling breaks down and the calculations from the hadronic transport model JAM reproduce the transverse momentum dependence of the measured $v_2(p_T)$, implying hadronic interaction dominance.
As collision energy increases, 
%a gradual restoration of NCQ scaling is observed, 
a gradual onset of NCQ scaling is observed
and the hadronic transport model underpredicts, while the multiphase transport model more accurately captures the collectivity observed in the 4.5 GeV data.
The observed breakdown and subsequent onset of NCQ scaling suggest 
a dominance of partonic interactions in collisions at {\rootsNN} $\ge$ 4.5 GeV, 
signaling the emergence of partonic collectivity and the possible formation of QGP.
These results shed light on the energy-dependent onset of QGP and 
contribute to the broader effort to map the QCD phase diagram.

%%%%%%%%%%%%%%%%%%  Main article  %%%%%%%%%%%%%%%%%%

%\section*{Acknowledgments}
%\vspace{10cm}
{\itshape Acknowledgments---}We thank the RHIC Operations Group and SDCC at BNL, the NERSC Center at LBNL, and the Open Science Grid consortium for providing resources and support.  This work was supported in part by the Office of Nuclear Physics within the U.S. DOE Office of Science, the U.S. National Science Foundation, the Ministry of Science and Technology of China and the Chinese Ministry of Education, National Natural Science Foundation of China, Chinese Academy of Science, NSTC Taipei, the National Research Foundation of Korea, Czech Science Foundation and Ministry of Education, Youth and Sports of the Czech Republic, Hungarian National Research, Development and Innovation Office, New National Excellency Programme of the Hungarian Ministry of Human Capacities, Department of Atomic Energy and Department of Science and Technology of the Government of India, the National Science Centre and WUT ID-UB of Poland, the Ministry of Science, Education and Sports of the Republic of Croatia, German Bundesministerium f\"ur Bildung, Wissenschaft, Forschung and Technologie (BMBF), Helmholtz Association, Ministry of Education, Culture, Sports, Science, and Technology (MEXT), and Japan Society for the Promotion of Science (JSPS).

{\itshape Data Availability---}The data that support the findings of this article are openly available~\cite{HEPData:159489}.

\bibliography{example} % Produces the bibliography via BibTeX.

\end{document}